\documentclass[aps,prl,showpacs,preprint]{revtex4}

\usepackage{graphicx} 
\usepackage{amsmath}
\usepackage{wrapfig}

\begin{document}

\title{Subradiance and superradiance-to-subradiance transition in dilute atomic clouds}
\author{{Diptaranjan Das, B. Lemberger, and  D. D. Yavuz}}
\affiliation{Department of Physics, 1150 University Avenue,
University of Wisconsin at Madison, Madison, WI, 53706}
\date{\today}
\begin{abstract}

We experimentally study subradiance in a dilute cloud of ultracold rubidium (Rb) atoms where $n \lambda_a^3 \approx 10^{-2}$ ($n$: atomic density, $\lambda_a$ excitation wavelength) and the on-resonance optical depth of the cloud is of order unity. We show that in the strong excitation regime, the subradiant time-scales depend on the excitation fraction of the cloud; i.e., to the intensity of the excitation pulse. In this regime, the decay dynamics are highly complicated and there is not a single decay time-constant. Instead, the decay time constant varies during the dynamics. Specifically, we were able to observe signatures of superradiant-to-subradiant transition; i.e., initially the decay rate is faster than independent decay (superradiant emission), while at later times it transitions to slower (subradiant emission). We also discuss a theoretical model whose numerical results are in good agreement with the experiments. 

\end{abstract} 
\maketitle

\section{I. Introduction}

Since the seminal paper by Dicke \cite{dicke}, collective-decay (superradiant or subradiant) of an ensemble of radiators has been studied by many authors and this problem continues to be relevant for a wide range of physical systems \cite{haroche,yelin,francis,kurizki}. Much of the physics of collective decay can be understood from a classical viewpoint. If the radiation from the individual emitters interfere constructively, then the total radiated power is coherently enhanced, resulting in a decay rate larger than the independent decay rate of the individual emitters. For example, if $N$ emitters are localized to a spatial region much smaller than the wavelength of radiation, their emissions can all be in phase, producing a decay rate $\sim N$ times faster than independent decay, $\sim N \Gamma_a$. This is the well-known Dicke limit of superradiance, and has been extensively analyzed both theoretically and experimentally \cite{haroche}. In contrast, if the individual radiators are anti-phased appropriately, their radiation can interfere destructively, producing a decay rate slower than the independent rate (subradiance). In the Dicke limit, a simple approach to achieve subradiance would be to group the atoms in pairs, where atoms in each pair oscillate exactly out-of-phase. 

While superradiance was first experimentally observed more than four decades ago \cite{feld,manassah}, subradiance in an ensemble of atoms was experimentally demonstrated only very recently \cite{kaiser}. The reason for this is that to observe subradiance, appropriate out-of-phase superpositions of the emitters need to be maintained for time scales that are long compared to the independent decay time. As a result, the subradiant states are fragile and are quite susceptible to dephasing. To overcome this challenge, recent observation of subradiance utilized ultracold atomic clouds at low temperatures, thereby avoiding motional dephasing. We also note that collective effects are most pronounced when there are a large number of emitters within a wavelength cube of volume; i.e., in the Dicke limit as discussed above. However, collective effects remain and can be quite important well outside this limit, even when the average spacing between the emitters is larger than the wavelength. In fact, as pointed out in Ref.~\cite{kaiser}, it is easier to observe subradiance outside the Dicke limit, since Van der Waals dephasing due to short-range interactions is avoided \cite{haroche}. 

Recent observation of subradiance used a large ultracold cloud with a very high on-resonance optical depth \cite{kaiser}; an optical depth (OD) of 40 or higher. As a result, the interpretation of the data is complicated by the fact that radiation trapping also plays an important role \cite{kaiser_rt}. Furthermore, subradiance was observed in the weak excitation limit, where single atom excited subspace (which is of dimension $N$) is a reasonable approximation to the dynamics of the full Hilbert space. In this paper, we extend these pioneering results to the strong excitation regime, and also to much-more dilute ultracold clouds with an on-resonance OD of order unity. We show that in this regime, the subradiant time-scales depend on the excitation fraction of the cloud, which is determined by the intensity of the excitation pulse. We find that in this regime the decay dynamics are highly complicated and there is not a single decay time-constant. Instead, the decay time constant varies during the dynamics. Specifically we were able to observe signatures of superradiant-to-subradiant transition. At early times of the evolution the decay rate is faster than independent decay (superradiant emission), while at later times it transitions to slower rate (subradiant emission). 

The collective decay problem from large samples in the strong-field excitation regime is notoriously quite difficult, since the dimension of the Hilbert space is $2^N$, and there are no obvious symmetries that can be utilized to simplify the problem. There are very limited analytical and numerical tools that can be utilized in this regime \cite{haroche}. We discuss a theoretical formalism that produces numerical results which are in good agreement with our experimental observations. The formalism is motivated by the recently discovered full eigenvalue spectrum of the exchange Hamiltonian, which is an effective description of the fundamental interaction resulting in collective decay: the exchange of a photon \cite{lemberger}. 

In other important prior work: collective decay effects have been studied experimentally in a wide range of physical systems such as cold molecules \cite{McGuyer}, a system of two trapped ions \cite{brewer}, on multi-level transitions in hot Gallium atoms \cite{Pavol}, in cold atoms at the vicinity of a single mode nanofiber \cite{rolston}, and in planar metamaterial arrays \cite{zheludev}. Subradiant atomic momentum states were recently observed in a Bose-Einstein Condensate (BEC) \cite{Zimm}. Studies of superradiant emission have been carried out in cold atoms in the weak excitation limit \cite{havey, kaiser, bromley} as well as in diamond nanocrystals \cite{tomasvolz} and hybrid solid state devices \cite{mayer} where it is possible to study the system in the Dicke limit. Recently switching between superradiant and subradiant states was demonstrated in a 10-qubit superconducting circuit \cite{wang}. With regard to recent theoretical work, most of these studies have focused on the weak excitation limit where a macroscopic two level atomic ensemble absorbs a single photon \cite{scully,scully2,scully3,Svidzinsky,eberly,kaiser5, kaiser6,kaiser7,kaiser8,kaiser9,kaiser3,scully8,vetter,kaiser10,scully5}. Even though this restricts the problem to a small subspace of the total Hilbert space there are several interesting effects that can be explored, for example, directional emission \cite{scully,scully2}, photon localization \cite{kaiser10}, and collective Lamb shift \cite{scully3, Svidzinsky}. With subradiant states being analogous to decoherence free subspaces, exploitation of subradiant states and tuning between superradiant and subradiant states can have applications in quantum memory devices and quantum information processing \cite{scully4, smartsev}. This has inspired a lot of work in studying subradiance in artificial structures like atomic arrays and with modified environments as in a cavity \cite{woggon,zoller,garcia, jenkins2, ritsch3, adams1, adams2,jenkins, ritsch1,ritsch2, petrov}.  Other studies of cooperative emission include an analysis by the 'Polarium model' \cite{glauber}, a study of spatial profile of subradiance \cite{agarwal1}, emission characteristics of entangled sources \cite{agarwal2}, and a recent analysis of many atom emission by renormalized perturbation theory \cite{bachelard}.

\section{II. Experimental schematic}

\begin{figure}[ht]
\vspace{-0cm}
\includegraphics[width=15cm]{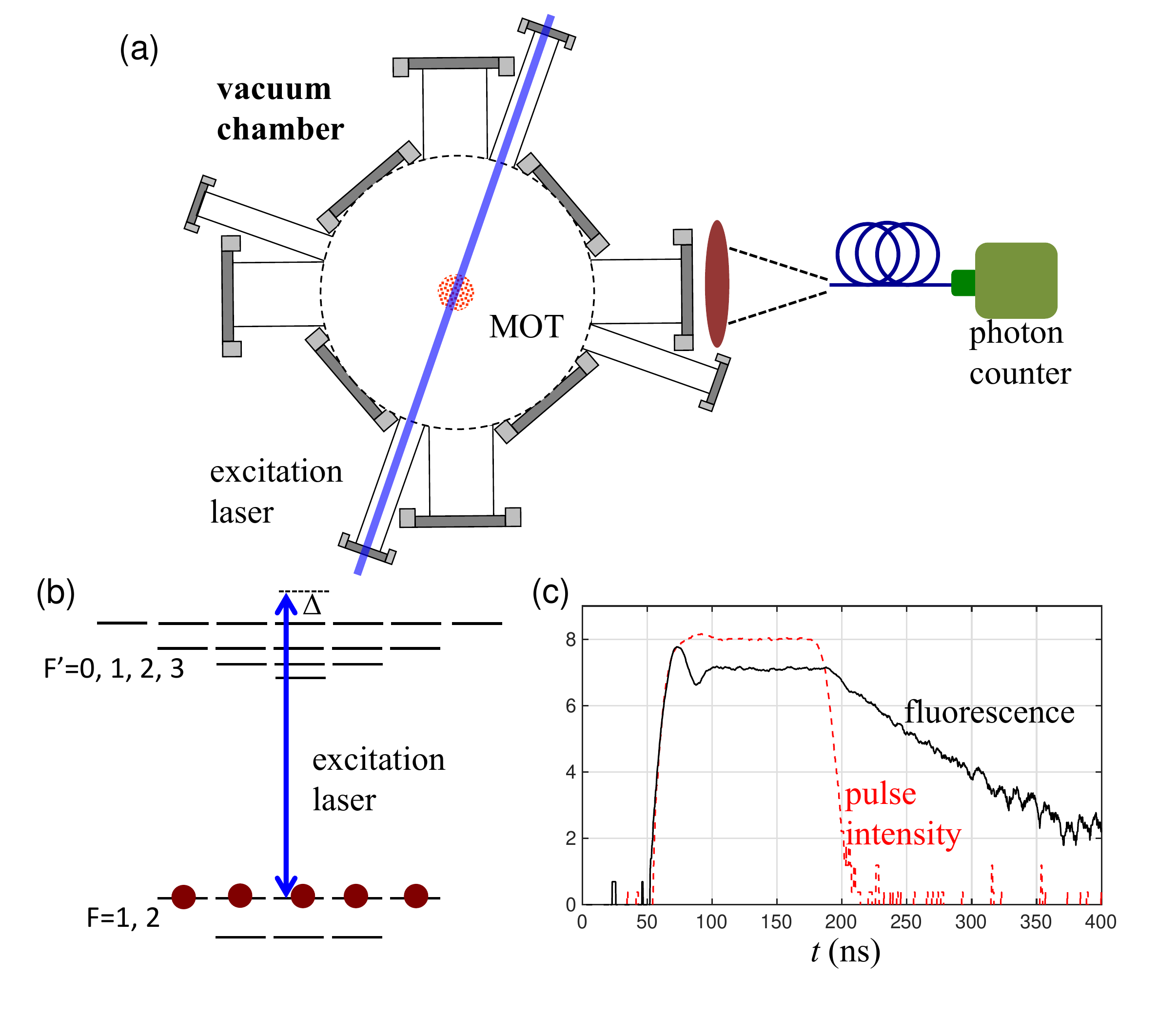}
\vspace{-0.4cm}
\caption{ (Color online) (a) The simplified experimental schematic. The experiment is performed inside a 14-port stainless-steel ultrahigh vacuum chamber. We first load the $^{87}$Rb atoms into a MOT from a background vapor. After the MOT is loaded and with the atoms optically pumped into the $F=2$ ground level, the atoms are excited into $F'=3$ level using a short and intense excitation laser. The fluorescence from the cloud is detected using a photon counter. (b) The relevant energy level diagram of $^{87}$Rb. (c) A sample fluorescence trace (solid black line) overlapped with an excitation pulse intensity trace (dashed red line), both plotted on logarithmic scale. }
\label{schematic}
\vspace{-0cm}
\end{figure}

We perform our experiments inside a 14-port stainless-steel ultrahigh vacuum chamber which is kept at a base pressure $5 \times 10^{-9}$~torr. A top view of our chamber is shown in Fig.~\ref{schematic}(a). We start the experiment by cooling and loading the atoms into a magneto-optical trap (MOT). To construct the $^{87}$Rb MOT, we use three counter-propagating beam pairs that are locked to the cycling $F=2 \rightarrow F'=3$ transition in the D2 line (transition wavelength of $\lambda_a=780$~nm), each with a beam power of about 50 mW and a beam size of 3 cm. The MOT lasers are obtained from a custom-built external-cavity diode laser (ECDL) whose output is amplified by semiconductor tapered amplifiers. Further details regarding our laser system can be found in our recent publications \cite{nickexp,jaredexp,brettexp}. The MOT lasers are overlapped with a hyperfine repumping beam, which is obtained from a separate ECDL locked to the $F=2 \rightarrow F'=2$ transition, with an optical power of about 1~mW.  

We load the atoms to the MOT from the background vapor for about 400~ms. In the last 40~ms of loading, we detune the MOT lasers to about $8 \Gamma_a$ ($\Gamma_a=2 \pi \times 6.02$~MHz is the decay rate of the transition) from the cycling transition, and reduce their intensity by about an order of magnitude to achieve efficient sub-Doppler cooling. At the end of the MOT loading cycle, we typically trap $\sim$~1.3~million atoms, within a radius of R=0.26~mm, giving an on-resonance OD of, OD~$=3N/(\kappa_a R)^2 \sim 1$ ($\kappa_a=2\pi/\lambda_a$ is the wave number at the transition wavelength). The atomic temperature is about 44~$\mu$K which is measured by monitoring the free-expansion of the cloud using an electron-multiplying CCD (EMCCD) camera. During the final 10~ms of the MOT loading cycle, we turn-off the hyperfine repumper beam. As a result, the atoms are optically pumped into the $F=2$ ground level at the end of the cycle.

The relevant energy level diagram of $^{87}$Rb is shown in Fig.~\ref{schematic}(b). With the atoms optically pumped into the $F=2$ hyperfine ground level, we turn-off the MOT beams, and turn on a single short and intense laser that couples the atoms to the $F'=3$ excited level. This laser is termed the excitation laser and its duration is about 120~ns. With the atoms excited into the $F'=3$ level, we turn-off the excitation beam abruptly, and record the fluorescence from the atoms using a single-photon counting module. The fast switching of the excitation beam is achieved using an acousto-optic modulator (AOM). The 90\%-10\% turn-off time of the excitation laser is 8~ns. We accomplish such fast switching by careful adjustment of the beam size inside the AOM. 

For each photon detected, the photon-counter produces $\sim10$-ns-long electronic TTL pulse, which is then measured by a fast-sampling digital oscilloscope. To avoid saturation of the photon-counter, we limit the number of detected photons for each experimental cycle to mostly around a photon. As a result, the experimental cycle (MOT loading-optical pumping-excitation-fluorescence detection) needs to be repeated many times to obtain a trace with a good signal-to-noise ratio. We typically repeat the experimental cycle $\sim 20,000$ times to obtain a fluorescence trace. With each experimental cycle lasting for about 1~s, a fluorescence trace takes about 6 hours to record in the lab. A sample fluorescence trace overlapped with an excitation pulse intensity (both on a logarithmic scale) is shown in Fig.~\ref{schematic}(c). 

\section{III. Experimental data analysis}

The excitation pulse is detected on a fast photodiode with a bandwidth of 150~MHz after the chamber. We start the data analysis after the excitation pulse has been turned off, $t=0$  is defined to be the point where the pulse intensity has dropped to less than 10\% of its peak intensity. The pulse intensity detected on the photodiode typically becomes indistinguishable from background within five nanoseconds after this point. 

The fluorescence signal, which is recorded using the photon counter, is proportional to the optical power emitted from the cloud, and we denote this signal by $P(t)$. Most collective-decay analysis, including Dicke's original paper, focuses on the total amount of excitation in the ensemble (i.e., the population of the excited level), which is proportional to the energy stored in the cloud. We denote this quantity by $E(t)$, which is related to the emitted power through the relation $E(t)=E_0-\int_0^t P(t')dt'$ (the inverse relationship is $P(t)=-dE(t)/dt$). Here, $E_0$ is the initial (at $t=0$) energy stored in the atomic cloud. Below, we will be plotting normalized versions of these quantities, redefined as $P(t) \equiv P(t)/P(t=0)$, and $E(t) \equiv E(t)/E(t=0)$. For independent decay, there is no difference between the time evolution of these two quantities since they have identical time dynamics: $E(t) \sim P(t) \sim \exp(-t/\tau_a)$ ($\tau_a$ is the lifetime of the excited level $\tau_a =1/\Gamma_a=26.2$~ns).

\begin{figure}[ht]
\vspace{-0cm}
\includegraphics[width=12cm]{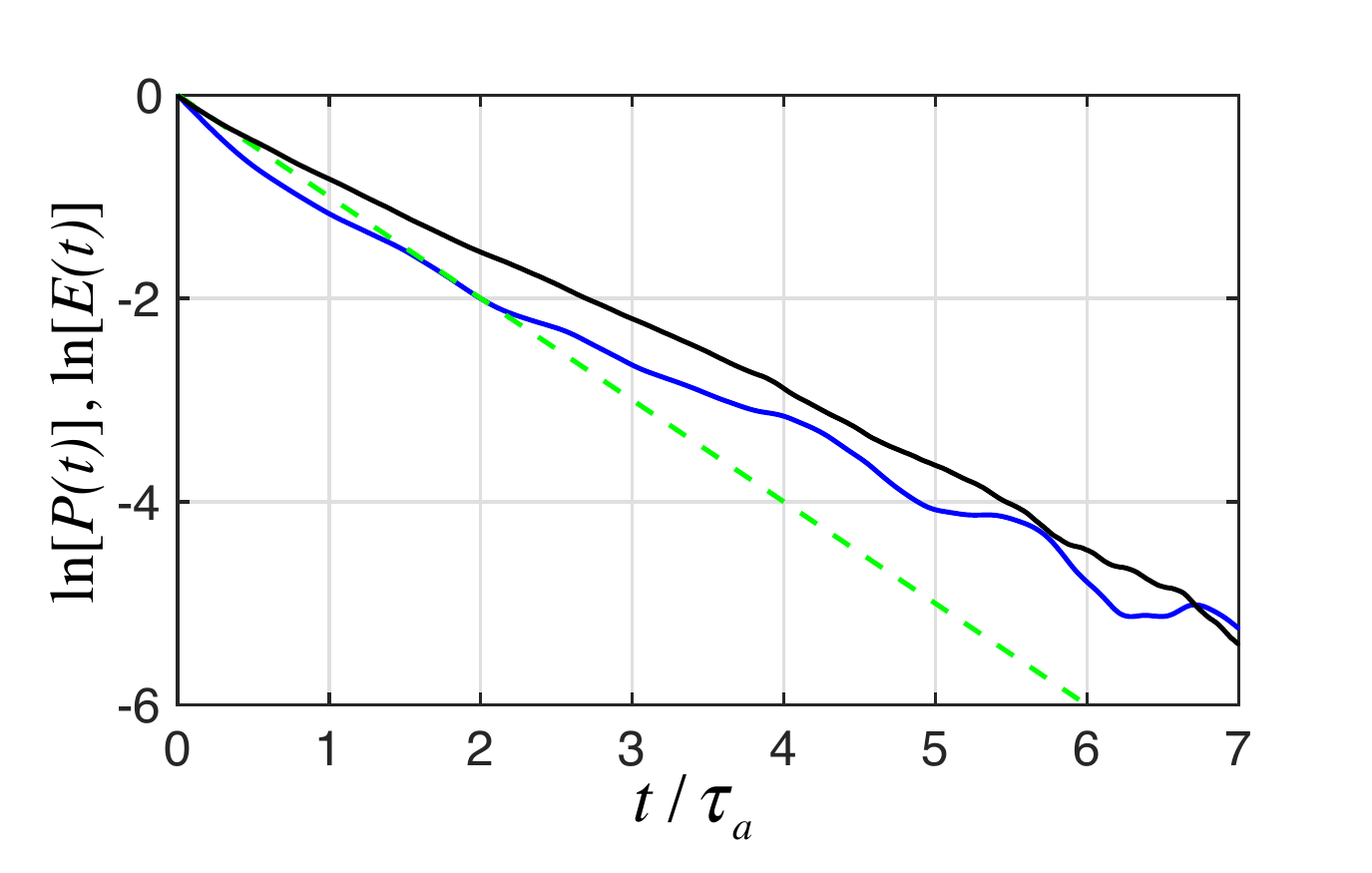}
\vspace{-0.4cm}
\caption{ (Color online) The observed fluorescence $P(t)$ (solid blue line), and the inferred stored energy in the cloud, $E(t)=E_0-\int_0^t P(t')dt'$ (solid black line) as a function of time for a sample dataset.  Both quantities are appropriately normalized and their natural logarithms are plotted (see text for details). For comparison, the case of independent decay, $\exp(-t/\tau_a)$, is also plotted (dashed green line).}
\label{power_vs_population}
\vspace{-0.2cm}
\end{figure}

Although the photon counter detects $P(t)$, we find instead that working with $E(t)$ is more convenient for most of the data discussed below. This is because $E(t)$ involves integration over the photon counter signal, which effectively amounts to averaging and reduces the noise.  An example of this is shown in Fig.~\ref{power_vs_population}, where we plot the natural logarithm of both of these quantities as a function of time, $\ln [P(t)]$ (solid blue line) and $\ln [(E(t)]$ (solid black line) for a sample dataset. For comparison, the case of independent decay $\exp(-t/\tau_a)$ is also plotted (dashed green line). The reduced noise in $\ln [(E(t)]$ can be clearly seen in the plots. Furthermore, the two curves do not lie on top of each other, which clearly shows that the decay is not a simple exponential decay, and cannot be described by a single decay time constant. As we discuss below, consistent with the theoretical model and the numerical results, the variation of the decay time constant during time evolution is better pronounced for $\ln [P(t)]$. The observed subradiance is quite remarkable considering that there are less than $10^{-2}$ atoms in a cubic wavelength of volume (i.e., $n \lambda_a^3 \approx 10^{-2}$) and the optical depth of the cloud is only of order unity. As we discuss below, the observed subradiance is consistent with the predictions of the theoretical model. 

\section{IV. Theoretical model and numerical simulations}

Superradiance and subradiance is well-known to be quite difficult to analyze in the large-sample and the strong excitation limit \cite{haroche}. In the Dicke limit, with all the atoms starting in the excited level, the system can be hypothesized to remain only in symmetric superpositions \cite{dicke}. This only leads to superradiance since symmetric superpositions are the states where the radiation from the emitters interfere constructively. For a large sample there are no obvious symmetries that can be employed and it is not clear how the exponentially large dimension of the Hilbert space can be simplified. Yet another complication is that for a sample which is spatially large compared to the wavelength, the phase of the emitted radiation varies between different emitters. 

\begin{figure}[ht]
\vspace{-0cm}
\includegraphics[width=6cm]{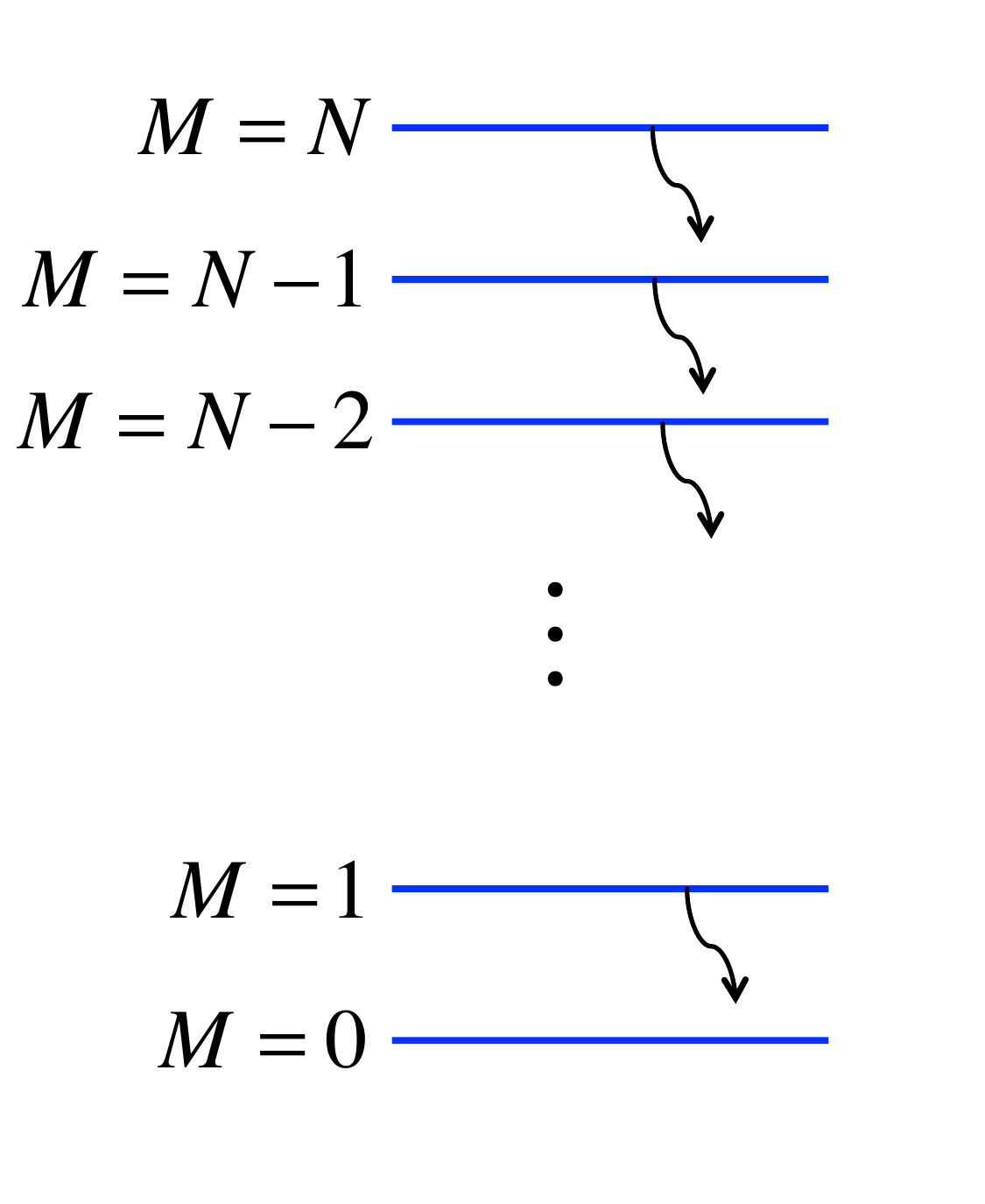}
\vspace{-0.4cm}
\caption{ (Color online) The excitation decay ladder for the formalism. Each subspace with $M$ atoms excited, decays to a subspace below (i.e., $M-1$ atoms excited).}
\label{decay_ladder}
\vspace{-0.2cm}
\end{figure}

To model collective-decay in the large sample and strong excitation regime, we extend the excitation ladder approach as discussed in Ref.~\cite{haroche}. The details of our formalism will be presented in the Appendix below, but we summarize the essential ideas here. It is well-known that  the excitation ladder approach quantitatively captures many aspects of Dicke superradiance \cite{haroche}. The key difficulty is how to extend this model for the large sample regime. For this purpose, we use a model that is motivated by the recently discovered eigenvalue spectrum of the exchange Hamiltonian, which is the basic physical interaction that causes correlated-decay \cite{lemberger}. 

\begin{figure}[ht]
\vspace{-0cm}
\includegraphics[width=13cm]{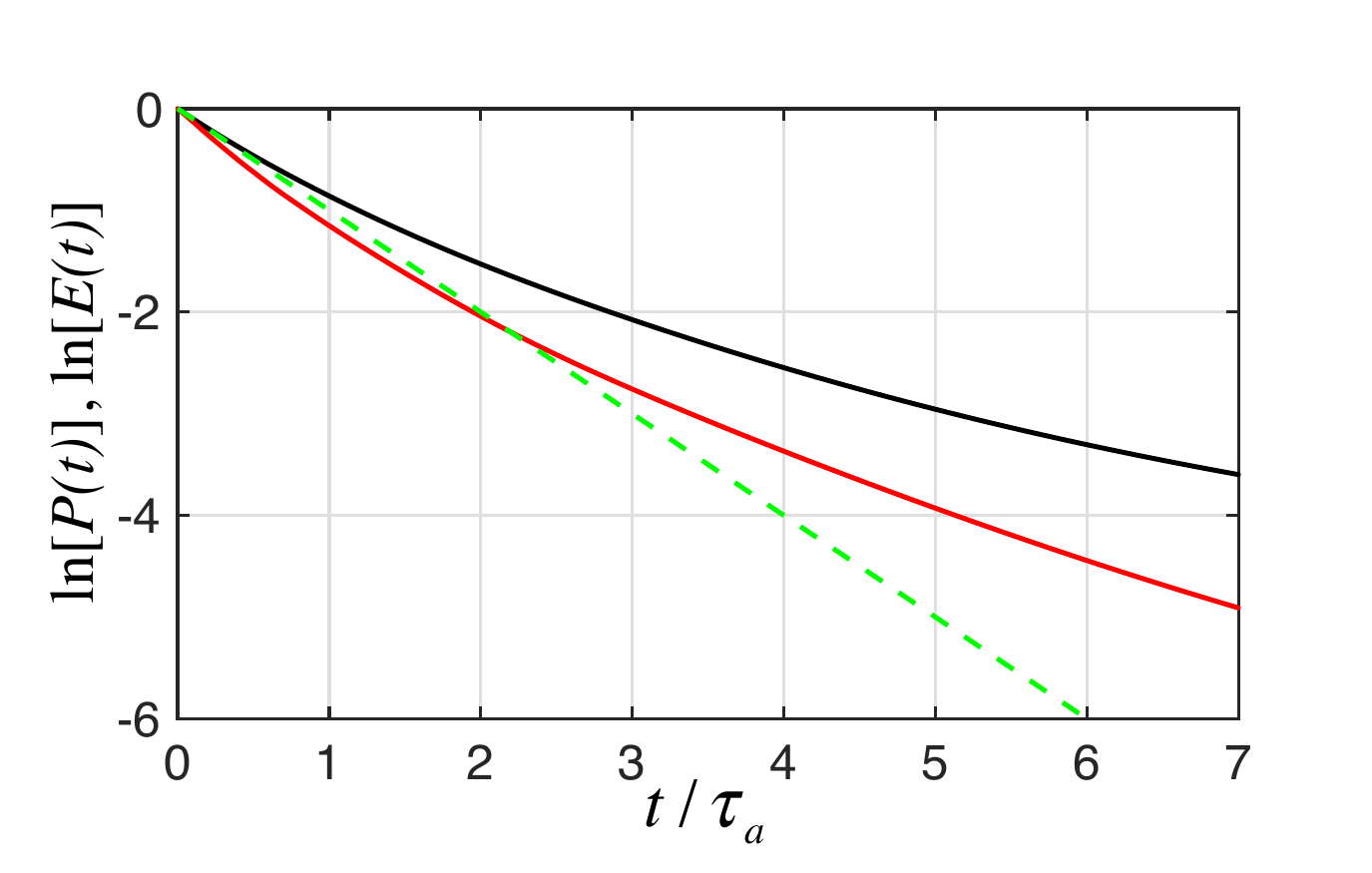}
\vspace{-0.4cm}
\caption{ (Color online) The stored energy $E(t)=\hbar \omega_a \sum_M M \rho_M(t)$ (solid black line) and the radiated power $P(t)=-dE(t)/dt$ (solid red line) for a numerical simulation for the nominal conditions of our experiment: $N=0.65$ million initially-excited atoms (1.3 million atoms with an excitation fraction of $0.5$) and a cloud radius of $R=0.26$~mm. For comparison, the case of independent decay, $\exp(-t/\tau_a)$, is also plotted (dashed green line). }
\label{numerical_super_to_sub}
\vspace{-0cm}
\end{figure}

We consider $N$ initially excited atoms uniformly distributed in a spherical cloud with a radius of $R$. The number of initially excited atoms is obtained by multiplying the total number of atoms with the excitation fraction.  We split the Hilbert space into subspaces that are indexed by $M=0, 1, ...., N$, which is the number of atoms in the excited state (while the remaining $N-M$ atoms are in the ground state). We denote the probability that the system is in $M$ atom excited subspace as $\rho_M(t)$. As shown in Fig.~\ref{decay_ladder}, each subspace $M$ decays to a subspace below (i.e., $M-1$ atoms excited). At $t=0$, the system starts in the $M=N$ subspace (i.e., at the top of the ladder), and then as time evolves decays down the ladder. We then have a coupled system of $N+1$ differential equations that describes the evolution of the system:
\begin{eqnarray}
\frac{d \rho_M}{dt} = -\Gamma_M \rho_M + \Gamma_{M+1} \rho_{M+1} \quad ,
\end{eqnarray}

\noindent where the quantity $\Gamma_M$ is the decay rate of subspace $M$ to subspace $M-1$. For independent (i.e., uncorrelated) decay, $\Gamma_M = M \Gamma_a$ since for independent decay the system wavefunction is a product of single-atom wavefunctions. The key idea of our formalism is that we modify this decay rate by a quantity which is proportional to the eigenvalue distribution of the exchange Hamiltonian for each subspace, with stimulated emission heuristically incorporated. Specifically, we take 
\begin{eqnarray}
\Gamma_M=M \Gamma_a + \xi \frac{\sqrt{\pi+29/12}}{k_a R} \sqrt{N-M} M \tilde{u} \Gamma_a \quad ,
\end{eqnarray}

\noindent where $\tilde{u}$ is a random variable whose value is uniformly distributed between $[-1,1]$. As shown in the Appendix, Eq.~(2) can be derived from our physical model with $\xi=1$. We use the dimensionless quantity $\xi$ as a free fitting parameter in the model, which can be viewed as the {\it shape} factor. This fitting parameter can be thought to account for (i) the deviation of the shape of the cloud from spherical, (ii) the uncertainty in the optical depth and, therefore, the atom number measurement of the cloud, and (iii) the uncertainty in the excitation fraction. As we discuss below, with this fitting parameter, this model successfully produces many aspects of our experimental results. In all the below fits, $\xi$ is of order unity and varies between $0.7-1$.  

Because the sign of the random variable $\tilde{u}$ can be positive or negative, each rate $\Gamma_M$ can be faster or slower than the independent decay case. For each simulation, we pick values for $\Gamma_M$ as given by Eq.~(2). With these values, we then numerically solve the $N+1$ coupled differential equations as given by Eq.~(1) using fourth order Runge-Kutta method, with the system starting at the top of the ladder (i.e., with the initial condition $\rho_N(t=0)=1$, and $\rho_M(t=0)=0$ for all $M \neq N$). For each simulation, we calculate the total energy stored in the cloud using $E(t)=\hbar \omega_a \sum_M M \rho_M(t)$. The radiated power is calculated using $P(t)=-dE(t)/dt$. To get an accurate description of the dynamics, we repeat the numerical simulation $\sim 1000$ times, picking different values for $\Gamma_M$ using Eq.~(2). We obtain the final result by averaging over these simulations.

Figure~\ref{numerical_super_to_sub} shows numerical results for our nominal experimental conditions: $N=0.65$ million initially-excited atoms ($1.3$ million atoms with an excitation fraction of $0.5$) and a cloud radius of $R=0.26$~mm. Here we plot the stored energy $E(t)$ (solid black) and radiated power $P(t)$ (solid red), both in logarithmic scale, as a function of time. Comparing Fig.~\ref{numerical_super_to_sub} to the experimental traces of Fig.~\ref{power_vs_population}, the model reasonably captures the overall subradiance, as well as the variation in the decay time scales. However, the model overestimates the change in the decay time-scales as the system evolves. One reason for this could be various dephasing mechanisms in the experiment, which is not accounted for in the model. 

\section{V. Experimental results}

\subsection{A. On-resonance versus detuned excitation}

\begin{figure}[ht]
\vspace{-0cm}
\includegraphics[width=18cm]{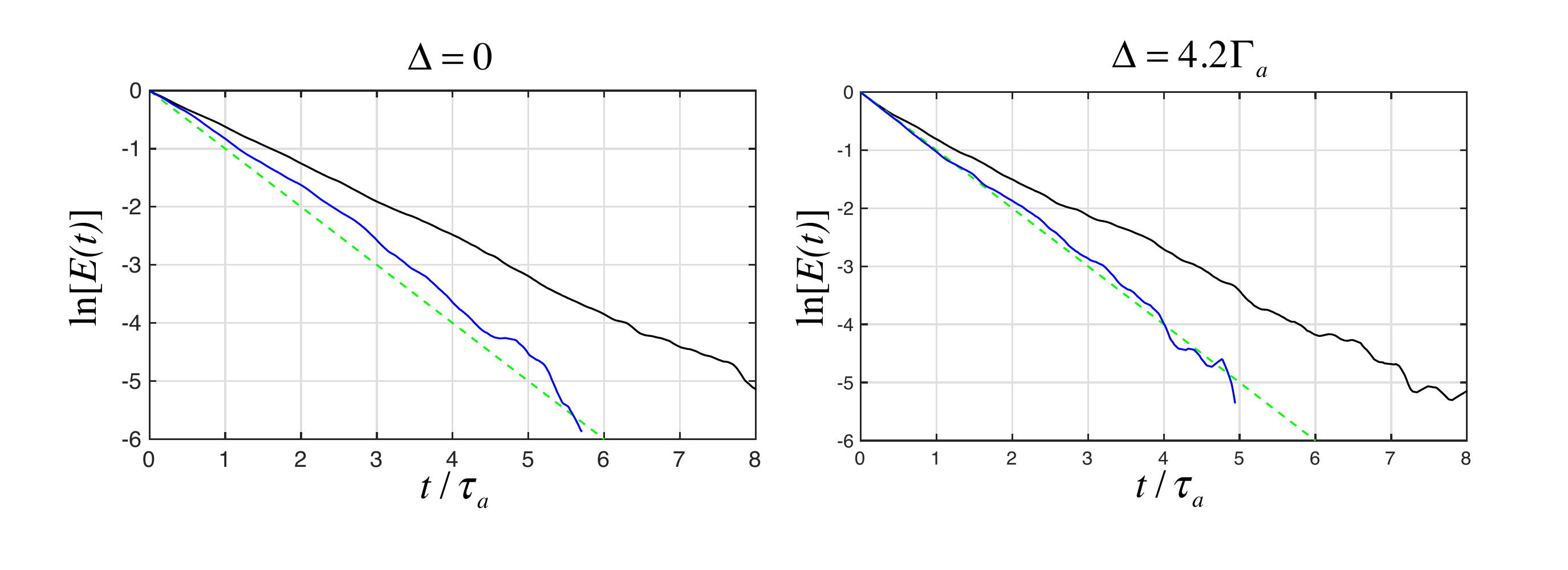}
\vspace{-1cm}
\caption{ (Color online)  Stored energy in the cloud (i.e., excited state population),  $\ln [(E(t)]$ for two different on-resonant optical depths, OD=1 (black line) and OD=0.35 (blue line). The plot on the left is obtained for an excitation laser which is on-resonant( $\Delta=0$), while for the plot on the right the excitation laser is detuned by an amount $\Delta=4.2 \Gamma_a$. For comparison, the case of independent decay, $\exp(-t/\tau_a)$, is also plotted (dashed green line). Because the results are qualitatively similar, the observed subradiance cannot be due to radiation trapping.  }
\label{on_off_resonance}
\vspace{-0cm}
\end{figure}

Because of the low on-resonant OD of our ultracold cloud, we do not expect incoherent photon absorption followed by reemission (i.e., radiation trapping) to play a role in our experiment. To experimentally confirm this, we compare fluorescence when the excitation laser is on-resonant versus off-resonant. Figure~\ref{on_off_resonance} shows $\ln [(E(t)]$ for two different optical depths, OD=1 (black line) and OD=0.35 (blue line), contrasting on-resonance ($\Delta =0$) versus detuned ($\Delta=4.2 \Gamma_a$) excitation. For both cases, the results are qualitatively similar with strong overall subradiance for OD=1, which shows that radiation trapping does not play a significant role. For a detuning of $\Delta=4.2 \Gamma_a$, the photon absorption probability (i.e., off-resonant optical depth) decreases by a factor of $ (2 \Delta /\Gamma_a)^2 \approx 70$ compared to its on-resonant value. As a result, if radiation trapping was responsible for the observed slower decay rate, there would be a large difference between on-resonance versus off-resonance excitation. 

All of the data shown in the rest of the paper is taken at a detuning of  $\Delta=4.2 \Gamma_a$.

\begin{figure}[h]
\begin{center}
\begin{minipage}[t]{0.5\linewidth}
\vspace{-2cm}
\includegraphics[width=10cm]{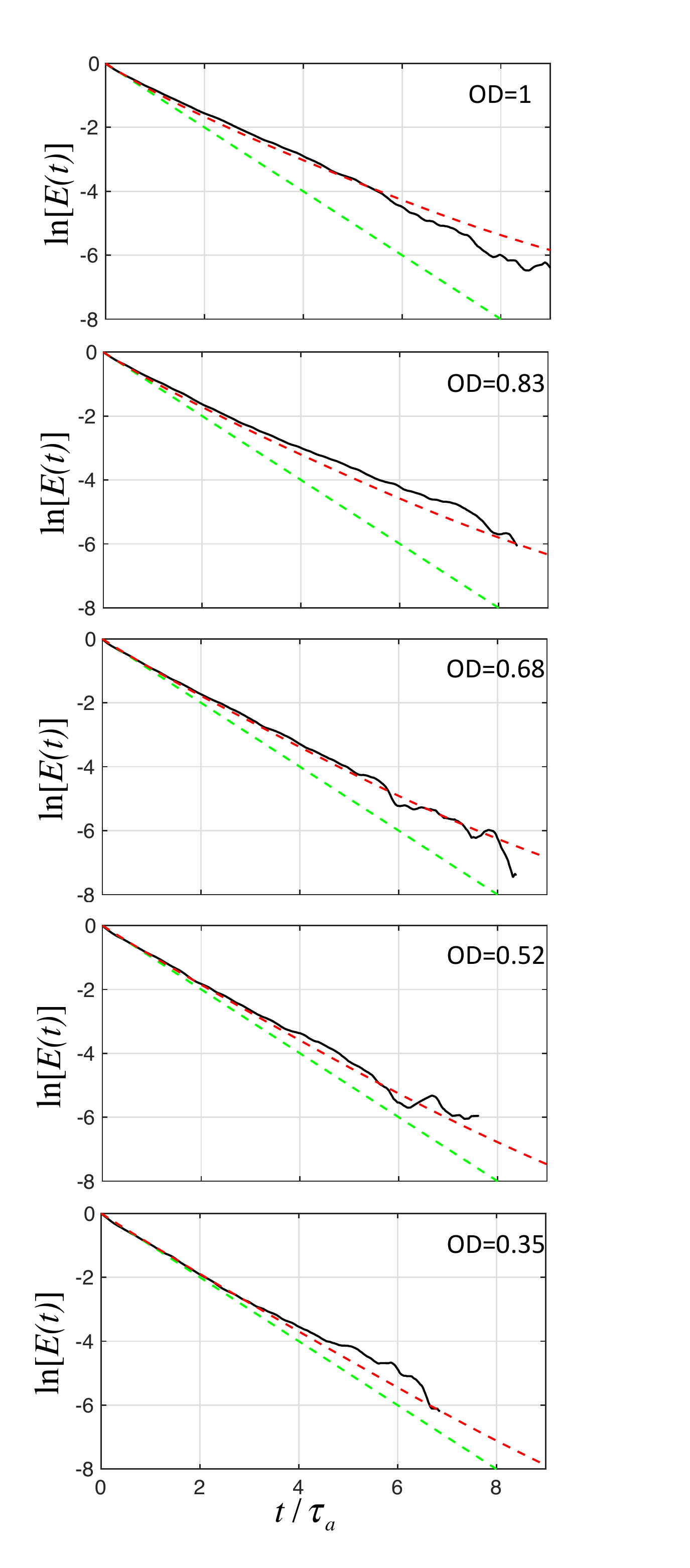}
\end{minipage} \hfill
\begin{minipage}[t]{0.45\linewidth}
\vspace{3cm} \caption{ (Color online) $\ln [(E(t)]$  as the optical depth is varied from OD=1 (top plot) to OD=0.35 (bottom plot). In each plot, the dashed red line is the result of the theoretical model discussed in the text.  For comparison, the case of independent decay, $\exp(-t/\tau_a)$, is also plotted (dashed green line). As expected, as the optical depth is reduced, the observed subradiance is reduced and the decay approaches to independent (i.e., uncorrelated) decay. Only one value of $\xi$ is used for all plots, demonstrating that the physical model can predict the dependence of collective effects on atomic density.  } 
\end{minipage}
\end{center}
\label{od_scan}
\vspace{-1cm}
\end{figure}

 \subsection{B. Optical depth scan}

Figure~6 shows $\ln [(E(t)]$  for an on-resonant optical depth of OD$=1, 0.83, 0.68, 0.52,$ and $0.35$, respectively. The optical depth is varied by turning off the MOT beams and letting the cloud free expand for a certain duration of time before the excitation beam is applied. The five optical depths are obtained after an expansion time of $0, 1, 2, 3,$ and $4$ ms, respectively. The optical depth after each expansion is calculated by measuring the size of the cloud using the EMCCD. As expected, as the optical depth is reduced, the subradiance is less pronounced and the decay rate approaches to that of independent (i.e., uncorrelated) decay. 

In each plot, the dashed red line is the result of the theoretical model with the free parameter adjusted to be $\xi = 0.77$. This parameter is adjusted once to get a good overall fit for $ 0<t< 9 \tau_a$ for the top plot (i.e. for OD=1). There is no further adjustment for the consequent plots. With this single fitting parameter,  there is good agreement between the experimental data and the numerical results. For comparison, the case of independent decay, $\exp(-t/\tau_a)$, is also plotted (dashed green line). 

\begin{figure}[ht]
\vspace{-0cm}
\includegraphics[width=13cm]{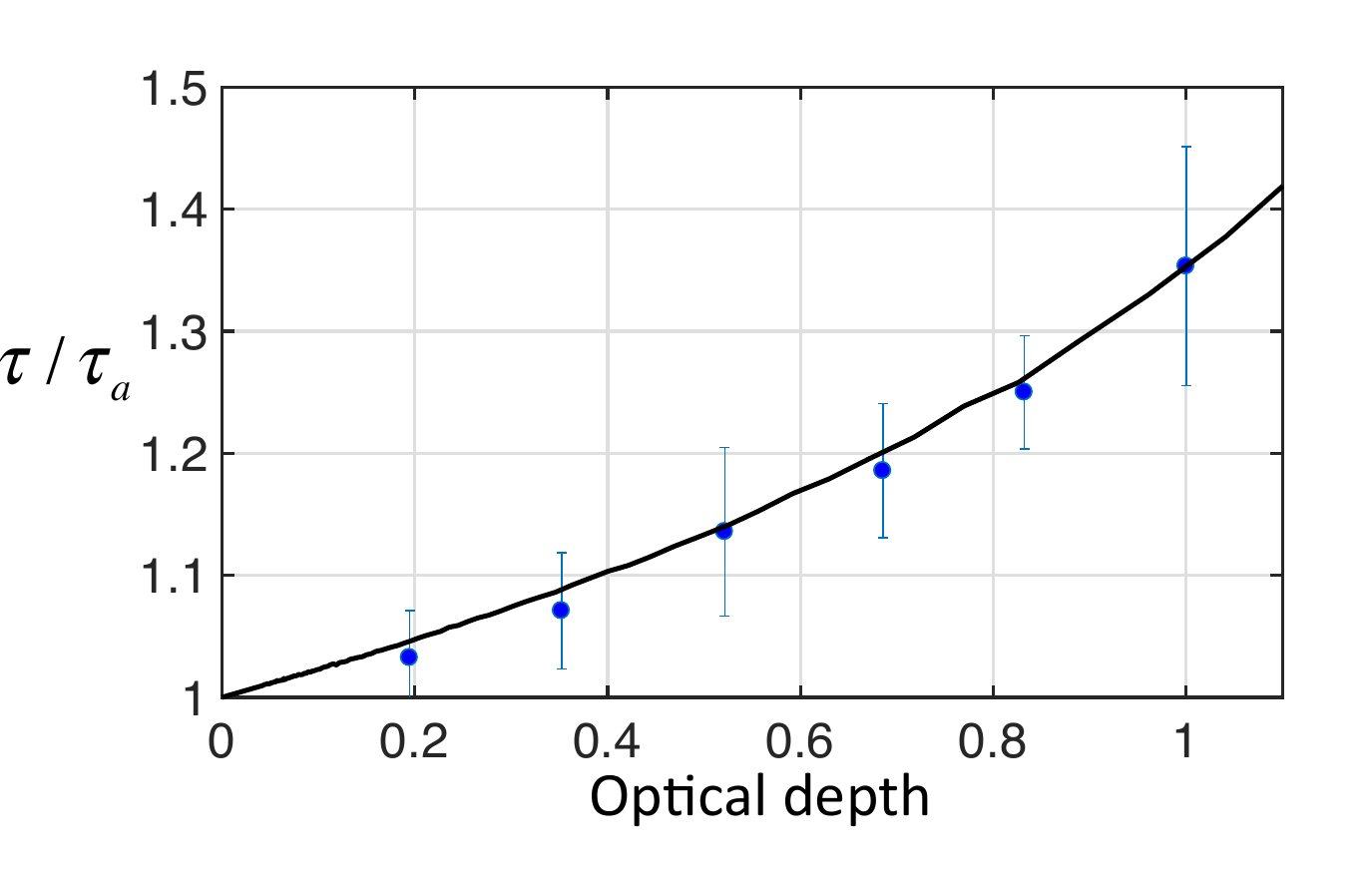}
\vspace{-0.4cm}
\caption{ (Color online) The mean decay time for each experimental curve shown in Fig.~6 during $0<t<$60~ns. The error bar in each data point is the standard variation of the decay time during this time window, and is therefore a measure of how much the decay time changes during the same time window. }
\label{od_decay_time}
\vspace{-0cm}
\end{figure}

For the data of Fig.~6, the decay is not a simple exponential decay and as a result there is not a single time constant. In Figure~\ref{od_decay_time}, we plot the mean decay time constant for each experimental curve shown in Fig.~6 during $0<t<2.3 \tau_a$. The error bar in each data point is the standard variation of the decay time during this time window, and is therefore a measure of how much the decay time changes during the same time window. The black curve is the result of numerical simulations where the free parameter is adjusted to get a good agreement for OD=1, $\xi = 0.94$. Again, with this single fitting parameter, there is good agreement between the experimental data and the numerical results. Consistent with the numerical results, there is some indication of a nonlinear dependence to the optical-depth, since the data points do not lie on a single line and instead curve upwards as the OD is increased.  

 \subsection{C. Excitation fraction scan}

\begin{figure}[ht]
\vspace{-0cm}
\includegraphics[width=13cm]{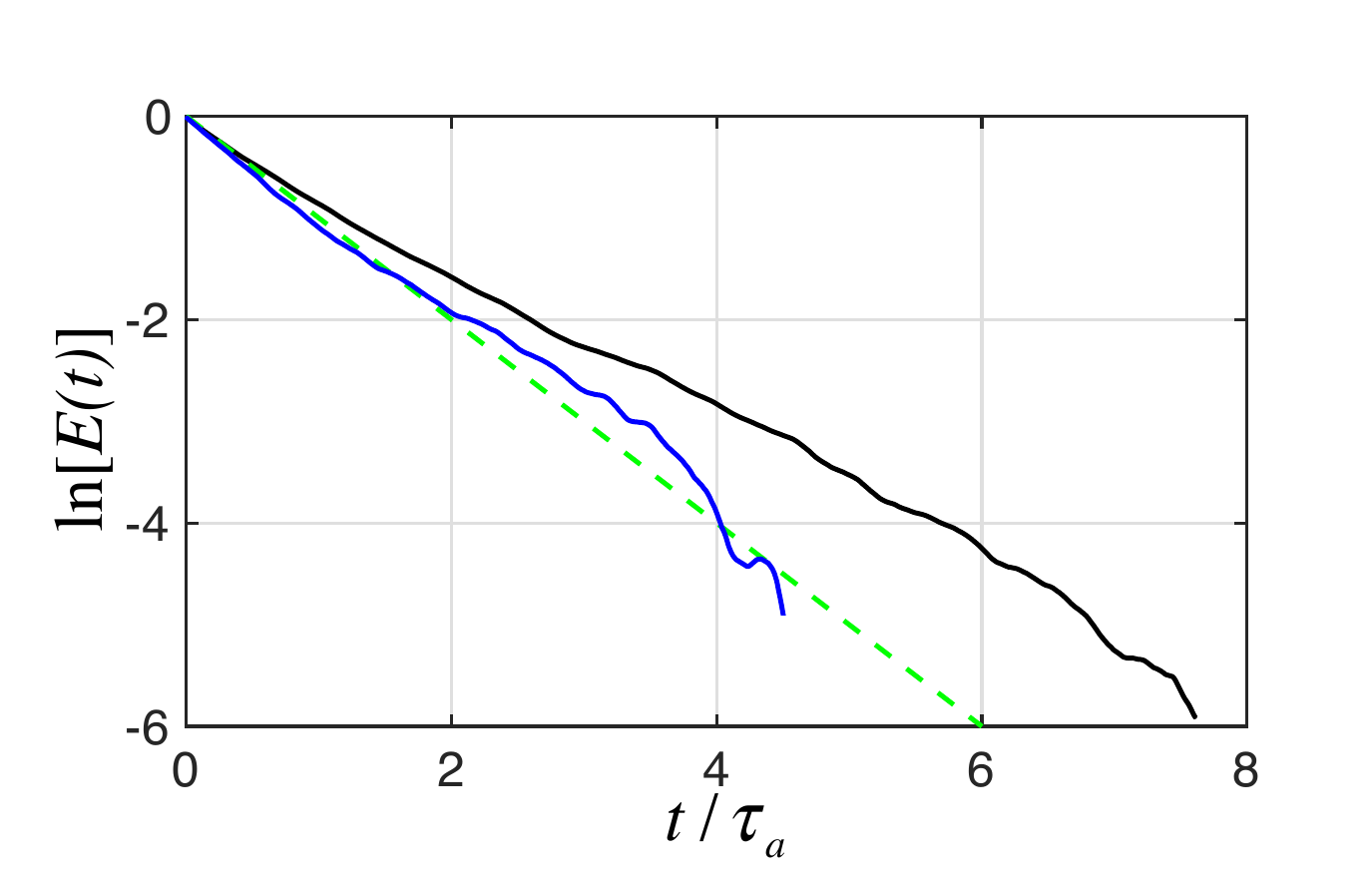}
\vspace{-0.4cm}
\caption{ (Color online)  $\ln [(E(t)]$ for high excitation fraction of 0.3 (solid black curve) and 0.08 (blue curve). For comparison, the case of independent decay, $\exp(-t/\tau_a)$, is also plotted (dashed green line). The amount of observed subradiance is significantly reduced as the excitation fraction is reduced. }
\label{int_scan}
\vspace{-0.2cm}
\end{figure}

In Ref.~\cite{kaiser}, subradiance was studied in the weak excitation regime where the single-atom excited subspace is a good approximation to the full dynamics. In this regime, the observed subradiant time-scales are independent of the intensity of the excitation laser. In this section, we discuss that in the dilute clouds and in the strong excitation regime, this is no longer the case. Figure~\ref{int_scan} shows $\ln [(E(t)]$ for high excitation fraction of 0.3 (solid black curve) and a relatively low excitation fraction of 0.08 (blue curve). For low excitation fraction, the decay approaches that of independent decay (dashed green line), and the observed subradiance is greatly reduced. 

\begin{figure}[ht]
\vspace{-0cm}
\includegraphics[width=13cm]{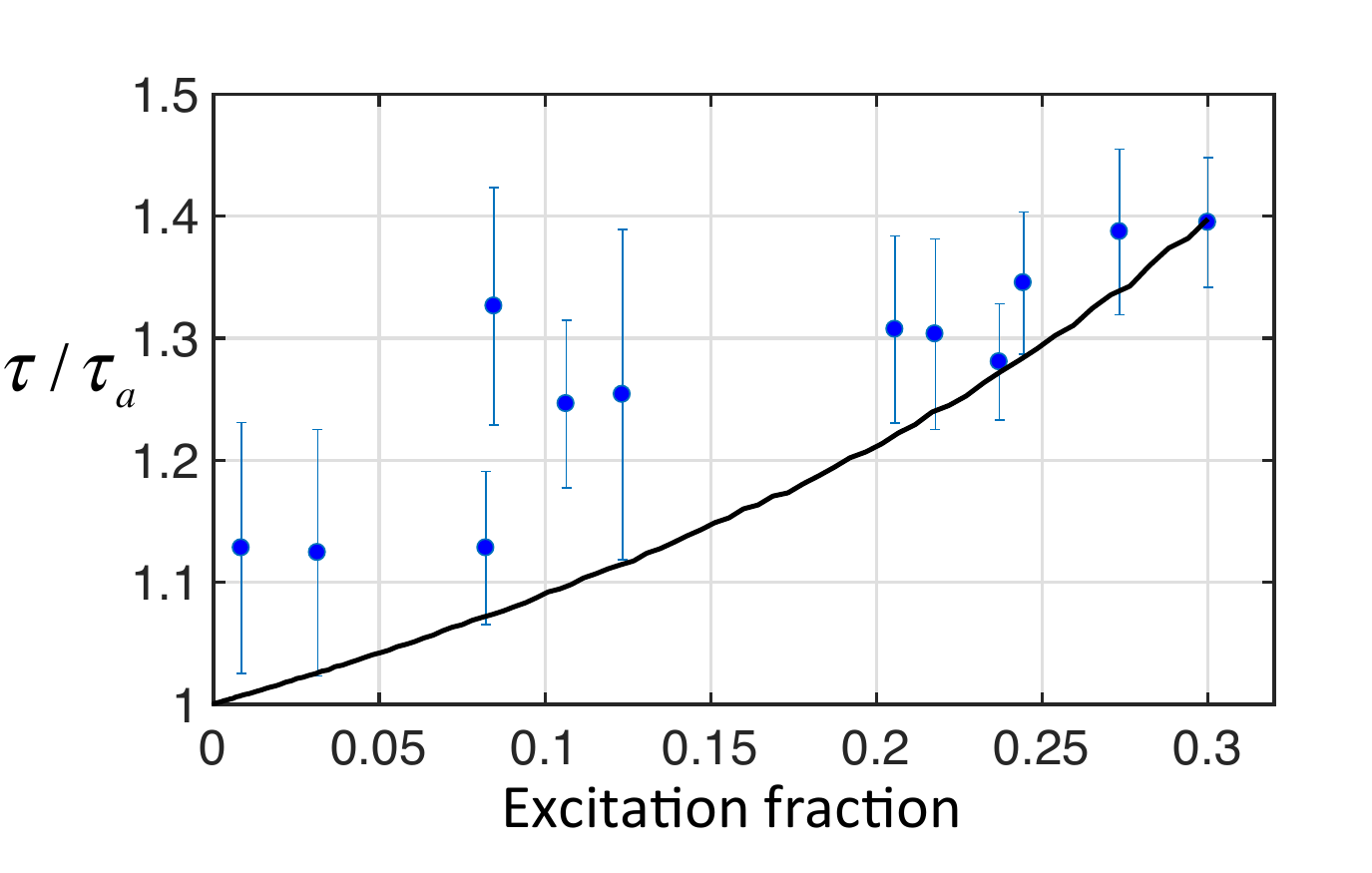}
\vspace{-0.4cm}
\caption{ (Color online) The mean decay time for each experimental curve shown in Fig.~6 during $0<t<2.3 \tau_a$. The error bar in each data point is the standard variation of the decay time, and is therefore a measure of how much the decay time changes during the same time window. The solid black curve is the result of numerical results with a single adjustable parameter. }
\label{int_decay_time}
\vspace{-0cm}
\end{figure}

Figure~\ref{int_decay_time} shows the mean decay time during $0<t<2.3 \tau_a$ for 12 experimental curves similar to the ones shown in Fig.~\ref{int_scan}. While there is a large spread in the data, there is also a clear trend that as the excitation fraction is increased, the decay time scales increase (i.e., the system becomes more subradiant). The solid black curve is the result of numerical results where the free parameter is adjusted to get a good agreement for the high excitation fraction of 0.3, $\xi=0.90$.  Again with this single fitting parameter, there is good agreement between the model and the experimental results.

 \subsection{D. Signatures of superradiance-to-subradiance transition}

In this section, we focus on the variation of the decay time as the system evolves; specifically on the superradiant-to-subradiant transition. For this purpose, we focus on the dynamics in the early times of the system evolution, $0<t<3 \tau_a$. As shown in Fig.~\ref{power_vs_population}, while both curves, in principle, contain the same amount of information, the variation of the decay time-constant during time evolution (i.e., how much each curve deviates from a linear line in the logarithmic plot) is more pronounced for the fluorescence curve. For this purpose, in this section we focus directly on the fluorescence as observed on the photon counter, $\ln [ P(t)]$. 

\begin{figure}[ht]
\vspace{-0cm}
\includegraphics[width=13cm]{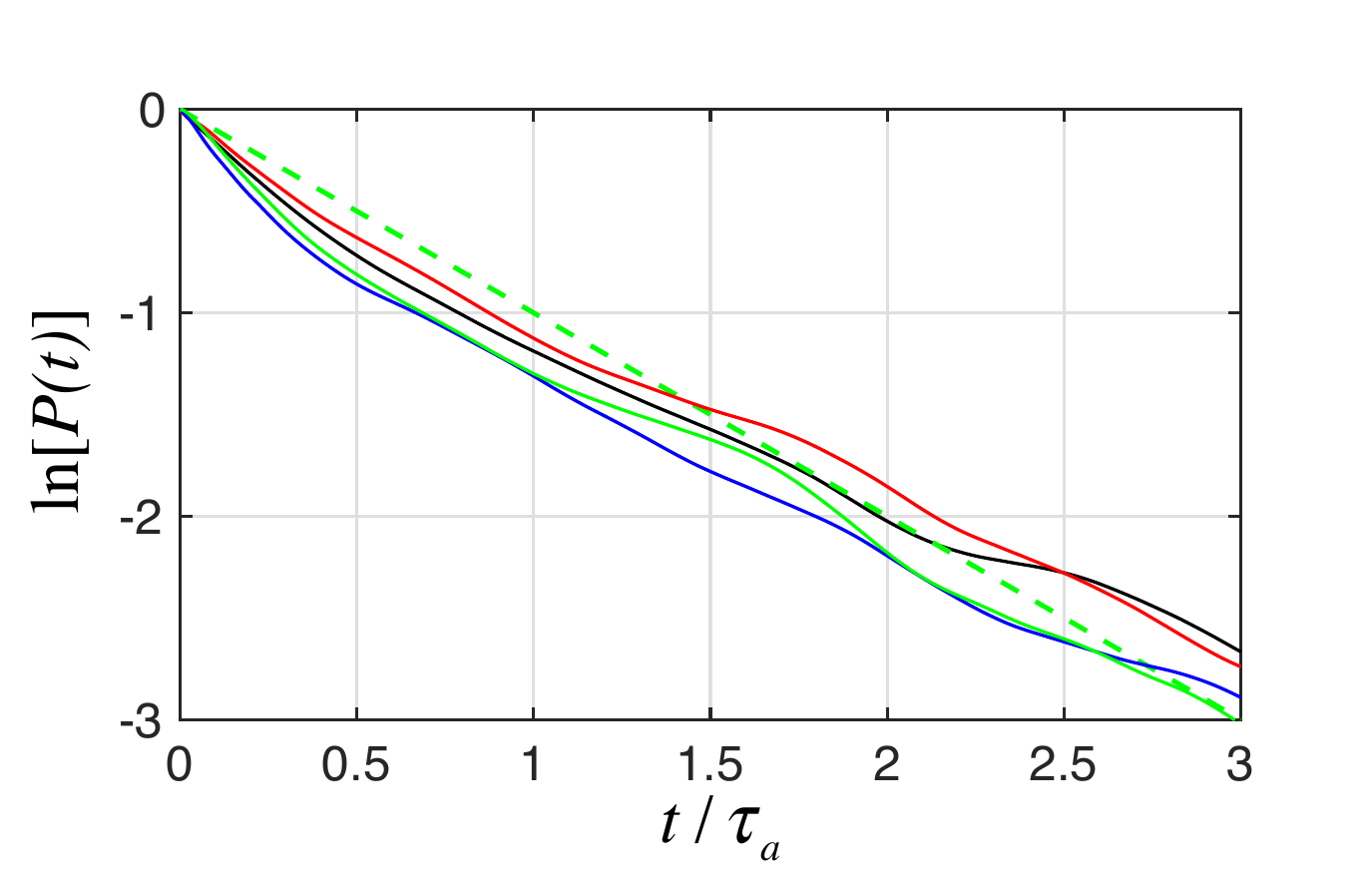}
\vspace{-0.4cm}
\caption{ (Color online) The observed fluorescence in logarithmic scale for a cloud optical depth of OD$=1$ (solid black), 0.83 (solid red), 0.68 (solid blue), and 0.52 (solid green). For comparison, the case of independent decay, $\exp(-t/\tau_a)$, is also plotted (dashed green line).}
\label{exp_super_to_sub}
\vspace{-0cm}
\end{figure}

Figure~\ref{exp_super_to_sub} shows the observed fluorescence in logarithmic scale, $\ln [ P(t)]$, for an on-resonant cloud optical depth of OD$=1$ (solid black), 0.83 (solid red), 0.68 (solid blue), and 0.52 (solid green). This data is obtained from the same data sets as the first four plots of Fig.~6. For comparison, the case of independent decay, $\exp(-t/\tau_a)$, is also plotted (dashed green line). For all the sets, faster than independent decay (superradiant) dynamics is evident for $t< \tau_a$. As the system evolves, this superradiance either evolves to subradiance (high optical depth; black and red curves), or to independent decay (low optical depth: blue and green curves). To our knowledge, this is the first observation of signatures of a superradiance-to-subradiance transition, which has been predicted by many theoretical papers \cite{haroche}. 

\begin{figure}[ht]
\vspace{-0cm}
\includegraphics[width=19cm]{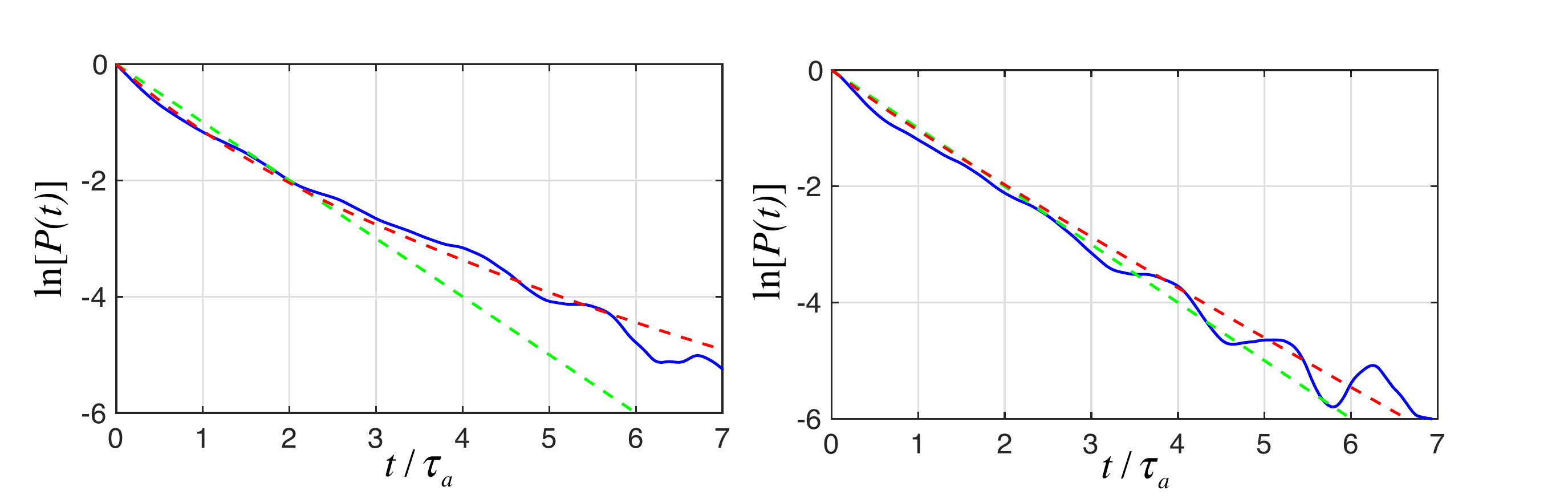}
\vspace{-0.4cm}
\caption{ (Color online) The observed fluorescence in logarithmic scale for a cloud optical depth of OD$=1$ (left) and 0.52 (right). For comparison, the numerical results are also plotted (dashed red curves).  }
\label{super_to_sub_exp_and_theory}
\vspace{-0cm}
\end{figure}

Figure~\ref{super_to_sub_exp_and_theory} shows the observed fluorescence for two optical depths OD$=1$ (high) and OD$=0.52$ (low) over a longer time window $0<t<7 \tau_a$, and also overlapped with the numerical results (dashed red curves). Here, the free parameter is adjusted only once to be $\xi=0.9$, in order to get good agreement with the experimental results for the high optical depth (left plot). For this case, the numerical results capture the variation of the decay time constant during time evolution, as well as superradiance-to-subradiance transition very well. For the lower optical depth (right plot), the agreement between the experimental data and the numerical results is worse. Specifically, the experimental curve continues to show signatures of superradiance at early times, while the numerical results do not. The reason for this discrepancy is currently an open question. We speculate that one reason for the discrepancy could be the assumption of a uniform cloud in the numerical simulations. In the experiment, the density of the MOT is unlikely to be uniform, due to the complicated three dimensional interference pattern produced by the six MOT laser beams. Due to this interference, there are likely localized regions with a higher density, which may be responsible for the persistent superradiant feature at early times, even at low optical depths. 

\section{VI. Conclusions and future work}

In conclusion, we experimentally studied subradiance in a dilute cloud of ultracold $^{87}$Rb atoms with densities very far away from the Dicke limit ($n \lambda_a^3 \sim 10^{-2}$),  and where the on-resonance optical depth of the cloud is of order unity. Although collective decay is an old and well-studied problem, our results are unique in a number of ways. Perhaps most importantly, we were able to observe signatures of superradiant-to-subradiant transition; i. e., initially the decay rate is faster than independent decay (superradiant emission), while at later times it transitions to slower rate (subradiant emission). Such a transition has long been predicted to be an important feature of collective decay, but has not been observed before. We also showed that in the regime that we study (dilute cloud in the strong excitation regime), the subradiant time-scales depend on the excitation fraction of the cloud (i.e., on the intensity of the excitation pulse).

We also discussed a theoretical model whose results are in good agreement with the experiments. The model relies on extension of the well-known decay ladder of the excitation, where the decay rate of each subspace is modified in accordance with the eigenvalue distribution of the exchange Hamiltonian. The model captures the observed (i) variation of the decay time constant with optical depth, (ii) variation of the decay time constant with the excitation fraction, and (iii) the subradiant-to-superradiant transition. However, the model overestimates the variation of the decay time as the system evolves: the curves shown in Fig.~\ref{numerical_super_to_sub} deviate more from linear compared to experimental curves. The model also does not capture the persistent superradiance at early times that we observe in the experiment, even at low optical depths. 

Extension of our results to mesoscopic ultracold clouds, with atom numbers in the range of 100-1000 would be very interesting. Such a mesoscopic system can be studied by loading the atoms to a far-off-resonant dipole trap, which is formed by focusing a detuned laser overlapping with the MOT. By moving one of the mirrors of the focusing optics, the beam size at the focus, and therefore the size of the trap, can be precisely controlled. This would allow independent control of the number ($N$) and density ($n$) of atoms in the trap. Such highly controlled mesoscopic systems will likely allow for better probing of many of the physics that we have explored in this paper, including the superradiance-to-subradiance transition.  

Our results have important implications for a number of research areas. Perhaps most important immediate application is to quantum information science. As mentioned above, subradiant states have gained renewed attention over the last decade since they are less susceptible to decoherence. Our work experimentally shows that such states can, in principle, be prepared even in the large-sample, very dilute limit. 

On a more fundamental note, Ref.~\cite{lemberger} discussed the implications of cooperative effects for scalability of quantum computers. Specifically, it was shown that noise due to collective decay produced errors in a quantum computer beyond the applicability of the threshold theorem, and therefore outside the current models of quantum error correction. The key reason for this is that  cooperative effects cannot be ignored, even when the average distance between the qubits is larger than the emission wavelength. Our experiment indeed shows that this is the case: even when on average there is only 0.01 atoms in a cubic wavelength of volume, cooperative effects can be quite important.

\section{VII. Acknowledgements}

We thank David Gold, Josh Karpel, Zach Buckholtz, and Shay Inbar for many helpful discussions. We also would like to thank Jared Miles for his experimental contributions at the early stages of this project.

\section{Appendix: Details of the theoretical model and numerical simulations}

\subsection{A. Formalism and the exchange interaction}

Consider $N$ two-level atoms, each with levels $|0\rangle$ and $|1\rangle$, in a three-dimensional geometry. We denote each individual atom with the index $j$ and consider a continuum of electromagnetic modes with annihilation and creation operators $\hat{a}_{\kappa \epsilon}$ and $\hat{a}_{\kappa \epsilon}^\dag$ respectively. These operators act on the mode of the field with wave-vector $\kappa$ and polarization $\epsilon$. The total Hamiltonian for the system when only the energy conserving terms are retained (under the rotating wave approximation) is \cite{lemberger}:
\begin{eqnarray}
\hat{H}_{total} & = & \sum_{j} \frac{1}{2}\hbar \omega_a \hat{\sigma}_z^j + \sum_{\kappa \epsilon}\hbar \nu_{\kappa \epsilon} \left( \hat{a}_{\kappa \epsilon}^\dag \hat{a}_{\kappa \epsilon}+\frac{1}{2} \right)   \quad \nonumber \\ 
& - & \sum_{j} \sum_{\kappa \epsilon}\hbar g_{\kappa \epsilon} \left[ \hat{a}_{k \epsilon} \exp{(i \vec{\kappa} \cdot \vec{r}_j )} \hat{\sigma}_+^j+\hat{a}_{\kappa \epsilon}^\dag \exp{(- i \vec{\kappa} \cdot \vec{r}_j ) }\hat{\sigma}_{-}^j \right] \quad ,
\label{m1}
\end{eqnarray}

\noindent where
\begin{eqnarray}
\hat{\sigma}_z^j & =& |1\rangle^j \hspace{0.1cm} {^j}\langle 1|-|0\rangle^j \hspace{0.1cm} {^j}\langle 0| \quad , \nonumber \\
\hat{\sigma}_+^j & =& |1\rangle^j \hspace{0.1cm} {^j}\langle 0| \quad , \nonumber \\
\hat{\sigma}_{-}^j & =& |0\rangle^j \hspace{0.1cm} {^j}\langle 1| \quad .
\label{m2}
\end{eqnarray}

\noindent In Eq.~(\ref{m1}), the first two terms describe the atoms and the electromagnetic modes in the absence of any interaction whereas the third term describes the coupling between the two systems.  $\vec{r}_j$ is the position of the $j$'th atom and the energies of the atom states $|0\rangle$ and $|1\rangle$ are taken to be $-\frac{1}{2} \hbar \omega_a$ and $\frac{1}{2} \hbar \omega_a$, respectively. The Dicke limit of the above equations is obtained when the total size of the sample is assumed to be small compared to the  $\kappa$-vector of the relevant modes, i.e., $\vec{\kappa} \cdot \vec{r}_j \rightarrow 0$. 

It is now well-understood that the key physical effect that describes many different aspects of collective decay, including superradiance and subradiance is the exchange interaction. Starting with the Hamiltonian of Eq.~(\ref{m1}), this interaction has been derived using a variety of approaches by a number of authors \cite{yamamoto,sokolov,kurizki,molmer}. One such derivation of the exchange interaction Hamiltonian is given in our earlier paper, Ref.~\cite{lemberger}, which we summarize here. The derivation uses assumptions that are similar to the traditional Wigner-Weisskopf theory of spontaneous decay \cite{yamamotobook}. Briefly, we take the initial atomic system to be an arbitrary superposition (in general entangled state) and assume initially zero excitation in each electromagnetic mode $\kappa\epsilon$. We then study the problem in the interaction picture and integrate out the probability amplitudes of the continuum states using the usual Born-Markov approximation. Using this approach, the end result is the following effective interaction Hamiltonian:
\begin{eqnarray}
\hat{H}_{eff} = \sum_j \sum_{k} \hat{H}^{jk} \quad .
\label{m3}
\end{eqnarray}

\noindent Here, the sum is over all pairs of qubits and operators $\hat{H}^{jk}$ act nontrivially only on the qubits with indices $j$ and $k$
\begin{eqnarray}
\hat{H}^{jk}= F_{jk}  \hat{\sigma}_+^j  \hat{\sigma}_-^{k} +F_{kj}  \hat{\sigma}_-^{i}  \hat{\sigma}_+^{j} \quad ,
\label{m4}
\end{eqnarray} 

\noindent which is essentially a ``spin" exchange interaction (mediated by photon modes) with coupling constants of $F_{jk}$:
\begin{eqnarray}
F_{jk}= F_{kj} &=& -(i \frac{\Gamma_a}{2} + \delta \omega_a)  \frac{3}{2}  \left[  (1-\cos^2\theta_{jk}) \frac{\sin \kappa_a r_{jk}}{\kappa_a r_{jk}} 
+ (1-3\cos^2\theta_{jk})  (\frac{\cos \kappa_a r_{jk}}{(\kappa_a r_{jk})^2} -\frac{\sin \kappa_a r_{jk}}{(\kappa_a r_{jk})^3} )   \right] \quad . 
\label{m5}
\end{eqnarray}

\noindent Here, $\Gamma_a$ is the single-atom decay rate and $\delta \omega_a$ is the single-atom Lamb shift of the qubit transition. $r_{jk}$ is the distance between the two atoms, and $\theta_{jk}$ is the angle between the atomic dipole moment vector and the separation vector $\vec{r}_{jk}$. The quantity $\kappa_a$ is the wave vector for the electromagnetic modes energy-resonant with the qubit transition: $\kappa_a = \omega_a /c $.

\subsection{B. The width of the eigenvalue distribution for $M$-subspace}

In this section, we discuss the width of the eigenvalue spectrum of the exchange Hamiltonian $\hat{H}_{eff} = \sum_{jk} \hat{H}^{jk}= \sum_{k}  F_{jk}  \hat{\sigma}_+^j  \hat{\sigma}_-^{k} +F_{kj}  \hat{\sigma}_-^{j}  \hat{\sigma}_+^{k}$ in the $N \rightarrow \infty$ limit, for the $M$-atom excited subspace. In this limit, the eigenvalues $\lambda$ of $\hat{H}_{eff}$ can be viewed as having a continuous distribution with probability density function $f_{\Lambda}(\lambda) \equiv P \{ \Lambda = \lambda \}$. The width of the probability density function can be evaluated by explicitly calculating the second-moment (variance) of the distribution $\sigma^{(2)} \equiv E[\Lambda^2] = \int f_{\Lambda}(\lambda) \lambda^2 d\lambda$, where $E[...]$ stands for the expected value. By definition, this second moment is:
\begin{eqnarray}
\sigma^{(2)} & = & E[\Lambda^2] =  \left( \begin{array}{c} N \\ M \end{array} \right)^{-1} \text{Trace} \big[ \big( \hat{H}_{eff} )^2 \big] \nonumber \quad ,  \\
& = & \left( \begin{array}{c} N \\ M \end{array} \right)^{-1} \sum_q  \langle q |  \left( \sum_{jk}  F_{jk}  \hat{\sigma}_+^j  \hat{\sigma}_-^{k} +F_{kj}  \hat{\sigma}_-^{j}  \hat{\sigma}_+^{k} \right)^2 | q \rangle \quad . 
\label{m7}
\end{eqnarray}

\noindent Here, the summation $q$ is over all the states in the $M$ atom excited subspace. By inspection, each term  $ \langle q |  \left( \sum_{jk}  F_{jk}  \hat{\sigma}_+^j  \hat{\sigma}_-^{k} +F_{kj}  \hat{\sigma}_-^{j}  \hat{\sigma}_+^{k} \right)^2 | q \rangle $ produces $(N-M) M$ contributions, each approppriately scaled with the the square of the relevant coupling constant, $F_{jk}^2$. In the $N \rightarrow \infty$ limit, the result is therefore:
\begin{eqnarray}
\sigma^{(2)} & = & (N-M) M E[F_{jk}^2] \quad , \nonumber \\
& = & \frac{ \pi + 29/12}{k_a^2 R^2} (N-M) M \Gamma_a^2 \quad . 
\label{m8}
\end{eqnarray}

\noindent Here, in the last step, we have used the expected value of the squares of the coupling constants in a three-dimensional geometry, $E[F_{jk}^2]$, as discussed in Ref.~\cite{lemberger}. The standard deviation (width) of the distribution is the square-root of the variance given in Eq.~(\ref{m8}):
\begin{eqnarray}
\sigma = \sqrt{\sigma^{(2)}} = \frac{ \sqrt{\pi + 29/12}}{k_a R} \sqrt{N-M} \sqrt{M} \Gamma_a \quad . 
\label{m9}
\end{eqnarray}

\noindent The distribution is symmetric around $\lambda=0$, which means that there are an equal number of superradiant and subradiant states. We have numerically checked that the results are insensitive to the precise shape of the distribution; rather, as expected, the width is critical. As a result, we choose a simple uniform distribution centered around $\lambda=0$, with a width given by Eq.~(\ref{m9}).

\subsection{C. Heuristic incorporation of stimulated emission}

The formalism described above assumes each photon mode to be unoccupied initially, and as a result, it does not incorporate stimulated emission in the decay process. In the small sample regime, an $M$-atom subspace has ``$M$" photons stored, and the spontaneous rates would at most be enhanced by ``$M$", as the system decays through the ladder. This is because, the stimulated emission rate for an $M$-photon state is a factor of $M$ larger than the spontaneous rate \cite{yamamotobook}. For a large sample, all emitted photons would not interfere constructively, but instead interfere with random phases. As a result, we hypothesize that one would expect $\sqrt{M}$ enhancement compared to the spontaneous rate for the large sample. We, therefore, multiply the width given by Eq.~(\ref{m9}) by a factor of $\sqrt{M}$ to heuristically incorporate for stimulated emission.

\end{document}